\documentclass[nofootinbib,twocolumn,showpacs,superscriptaddress,twoside,prx]{revtex4-1}
\usepackage{amsmath}
\usepackage{amssymb}
\usepackage{amsfonts}
\usepackage{enumerate}
\usepackage{mathrsfs}
\usepackage{graphicx}
\usepackage{epstopdf}
\usepackage{enumerate}
\usepackage{changepage}

\usepackage{lipsum}

\newtheorem{theorem}{Theorem}

\newtheorem{observation}{Observation}

\newtheorem{proposition}{Proposition}

\usepackage[
colorlinks,
linkcolor = blue,
citecolor = blue,
urlcolor = blue]{hyperref}
\def \qed {\hfill \vrule height7pt width 7pt depth 0pt}

\newcommand{\ket}[1]{| #1 \rangle}

\newcounter{lastnote}

\begin{document}

\title{   Local distinguishability based  genuinely quantum nonlocality without entanglement}

\author{Mao-Sheng Li}

\affiliation{Department of Physics, Southern University of Science and Technology, Shenzhen, 518055, China}
	\affiliation{ Department of Physics, University of Science and Technology of China, Hefei, 230026, China}
\author{Yan-Ling Wang}
\email{wangylmath@yahoo.com}
\affiliation{ School of Computer Science and Techonology, Dongguan University of Technology, Dongguan, 523808, China}
\author{Fei Shi}
\email[]{shifei@mail.ustc.edu.cn}
\affiliation{School of Cyber Security,
	University of Science and Technology of China, Hefei, 230026,  China.}
\author{Man-Hong Yung}
\email{yung@sustc.edu.cn}
\affiliation{Department of Physics, Southern University of Science and Technology, Shenzhen, 518055, China}
\affiliation{Institute for Quantum Science and Engineering, and Department of Physics,
Southern University of Science and Technology, Shenzhen, 518055, China}


\begin{abstract}
Recently, Halder \emph{et al.} [\href{https://journals.aps.org/prl/abstract/10.1103/PhysRevLett.122.040403}{Phys. Rev. Lett. \textbf{122}, 040403 (2019)}] proposed the concept strong nonlocality without entanglement: an orthogonal set of fully product states in multipartite quantum systems that is locally irreducible for every bipartition of the subsystems.   As the difficulty of the problem, most of the results are  restricted to tripartite systems.  Here we consider a weaker form of nonlocality called local distinguishability based  genuine   nonlocality. A set of  orthogonal multipartite quantum states is said to be genuinely nonlocal if it is locally indistinguishable for every bipartition of the subsystems.  In this work, we tend to study the latter form of nonlocality. First, we present an elegant set of product states in bipartite systems that is locally indistinguishable. After that,  based on a simple observation, we present a general   method to construct genuinely nonlocal sets of multipartite product states by using those  sets that are genuinely nonlocal but with less parties. As a consequence, we obtain that genuinely nonlocal sets of fully product states exist for all possible multipartite quantum systems.

\end{abstract}

\maketitle
\section{Introduction}

Quantum states discrimination plays a fundamental role in quantum information processing. It is well known that a set of quantum states can be perfectly distinguished by positive operation value measurement (POVM) if and only if these states are pairwise orthogonal \cite{nils}. In a multipartite setting, due to the physical obstacles, sometimes  we can not take a global measurement but only can use  local operations with classical communication (LOCC).  Bennett et al. \cite{Ben99} presented examples
of  orthogonal product states that are indistinguishable under LOCC  and   named such a phenomenon as  quantum nonlocality without entanglement.  The nonlocality here is in the sense that  there exists some quantum information that could be inferred from global measurement but cannot be read from local correlations of the subsystems.   A set  of orthogonal  states which is indistinguishable under LOCC is also   called  as being locally indistinguishable or   nonlocal. The local indistinguishability has been practically applied in quantum cryptography primitives such as   data hiding \cite{Terhal01,DiVincenzo02} and secret sharing \cite{Markham08,Rahaman15,WangJ17}.

Since the work of Bennett et al. \cite{Ben99}, the problem of local discrimination of quantum states has attracted much attention.   The maximally entangled states and  the  product states,  as being two extreme sets  among the pure states,    their local distinguishability is the most attractive.
    Here we present an incomplete list of the results about the local distinguishability of maximally entangled states  \cite{Gho01,Wal00,Wal02,Fan04,Nathanson05,Cohen07,Bandyopadhyay11,Li15,Fan07,Yu12,Cos13,Yu115,Wang19,Xiong19,Li20} and  product states \cite{Ben99,Ran04,Hor03,Ben99b,DiVincenzo03,Zhang14,Zhang15,Zhang16,Xu16b,Xu16m,Zhang16b,Wang15,Wang17,Feng09,
Yang13,Zhang17,Zhangj17,Halder18,Li18,Halder1909,Xu20a,Xu20b}.
 Another direction of related research is to study how much resource of  entanglement are needed in order to distinguish quantum states which are locally indistinguishable \cite{Cohen08,Bandyopadhyay16,Zhang16E,Bandyopadhyay18,Lilv19}.
Another important sets which are known to be  locally indistinguishable are those   unextendible product bases (UPB),   sets of incomplete
 orthonormal product states whose complementary space has no product states \cite{Ben99b,DiVincenzo03,Feng06,J14,CJ15}.

Recently, Halder \emph{et al.} \cite{Halder19} introduced a  stronger form of local indistinguishability, i.e.,  local irreducibility. A set of multipartite orthogonal quantum states is said to be locally irreducible if it is not possible to locally eliminate one or more states from the set while preserving orthogonality of the postmeasurement states. Under this setting, they  proposed the concept  strong nonlocality without entanglement. A set of orthogonal multipartite product states is called to be strong nonlocality  if it is locally irreducible for every bipartition of the systems.   They provided the first two examples of strongly nonlocal sets of product states in $\mathbb{C}^3\otimes\mathbb{C}^3 \otimes\mathbb{C}^3$ and $\mathbb{C}^4\otimes\mathbb{C}^4 \otimes\mathbb{C}^4$ and raised the questions of how  to extend their results to multipartite quantum systems and sets of unextendible product bases \cite{Halder19}. Quite recently, Zhang \emph{et al.} \cite{Zhang1906} extended the concept of strong nonlocality to more general settings. However, there are only  a few sets which have been proven to be strongly nonlocal \cite{Halder19,Zhang1906,Rout1909,Rout1910,Tian20,Shi20S}.   Most of the known results are in the tripartite quantum settings.    Here we propose a  form of nonlocality called genuine nonlocality whose nonlocality is lying between the local distinguishablity based nonlocality and the local irreducibility based strong nonlocality (the definition here is slightly different from that defined by Rout \emph{et al.} in Ref. \cite{Rout1909}). A set of  orthogonal multipartite quantum states is said to be genuinely nonlocal if it is locally indistinguishable for every bipartition of the systems. A nature question arises: are there genuinely nonlocal set of fully product states for any possible multipartite quantum systems?  In this paper, we tend to address this problem.

The rest of this article is organized as follows. In Sec. \ref{second}, we give some necessary notation, definitions and some basic result of local nonlocality of bipartite product basis.  In Sec. \ref{third}, we present a general method to obtaining genuinely nonlocal set of multipartite product states.    Finally, we draw a conclusion and present some interesting problems in     section \ref{fifth}.
	\vskip 8pt

\section{Locally indistinguishable set of bipartite product states}\label{second}

  For any integer $n\geq 2$,  we denote $U(n)$ to be the set of all unitary matrices of dimensional $n$. And throughout this paper, we use the following subset of unitary matrices
  $$ U_{FL}(n):=\{(h_{ij})_{i,j}^{n}\in U(n) |  h_{1k},h_{k1} \neq 0, k=1,\cdots,n   \}.$$
  That is, the set of $n$ dimensional unitary matrices whose elements on the first and last rows are all nonzero.

  Let $n\geq 3$ be an integer and $\mathcal{H}$ be a Hilbert space of dimensional $n$. Assume  that  $\mathcal{B}=(|1\rangle,\cdots, |n\rangle)$ is an $n$-tuple of vectors in $\mathcal{H}$  and these $n$ vectors are consisting of an orthonormal basis of $\mathcal{H}$. And we call $\mathcal{B}$ an ordered  orthonormal basis of $\mathcal{H}$. For any $H=(h_{jk})_{j,k=1}^{n-1}\in U_{FL}(n-1)$,  we define two operations on $\mathcal{H}$ with respect to $\mathcal{B}$
  {\small $$H_{\mathcal{B}}^{(U)}:=\displaystyle\sum_{j=1}^{n-1}\sum_{k=1}^{n-1} h_{jk} |j\rangle\langle k|, H_{\mathcal{B}}^{(D)}:=\displaystyle\sum_{j=1}^{n-1}\sum_{k=1}^{n-1} h_{jk} |j+1\rangle\langle k+1|.$$
  }
  That is, under the computational basis $\{|1\rangle,\cdots, |n\rangle\}$, their matrix representations are as follows
  $$
  \begin{array}{c}
  H_{\mathcal{B}}^{(U)}=\left[\begin{array}{ll}
  H& \mathbf{0}_{(n-1)\times 1}\\
  \mathbf{0}_{1\times (n-1)}& 0
  \end{array}
  \right],\\[3mm]
     H_{\mathcal{B}}^{(D)}=\left[\begin{array}{ll}
  0&  \mathbf{0}_{1\times (n-1)}\\
 \mathbf{0}_{(n-1)\times 1}& H
  \end{array}
  \right].
  \end{array}
   $$
We call them the up  extension and down extension of $H$ with respect to   $\mathcal{B}$ respectively.

	\begin{figure}[h]
		\includegraphics[width=0.48\textwidth,height=0.38\textwidth]{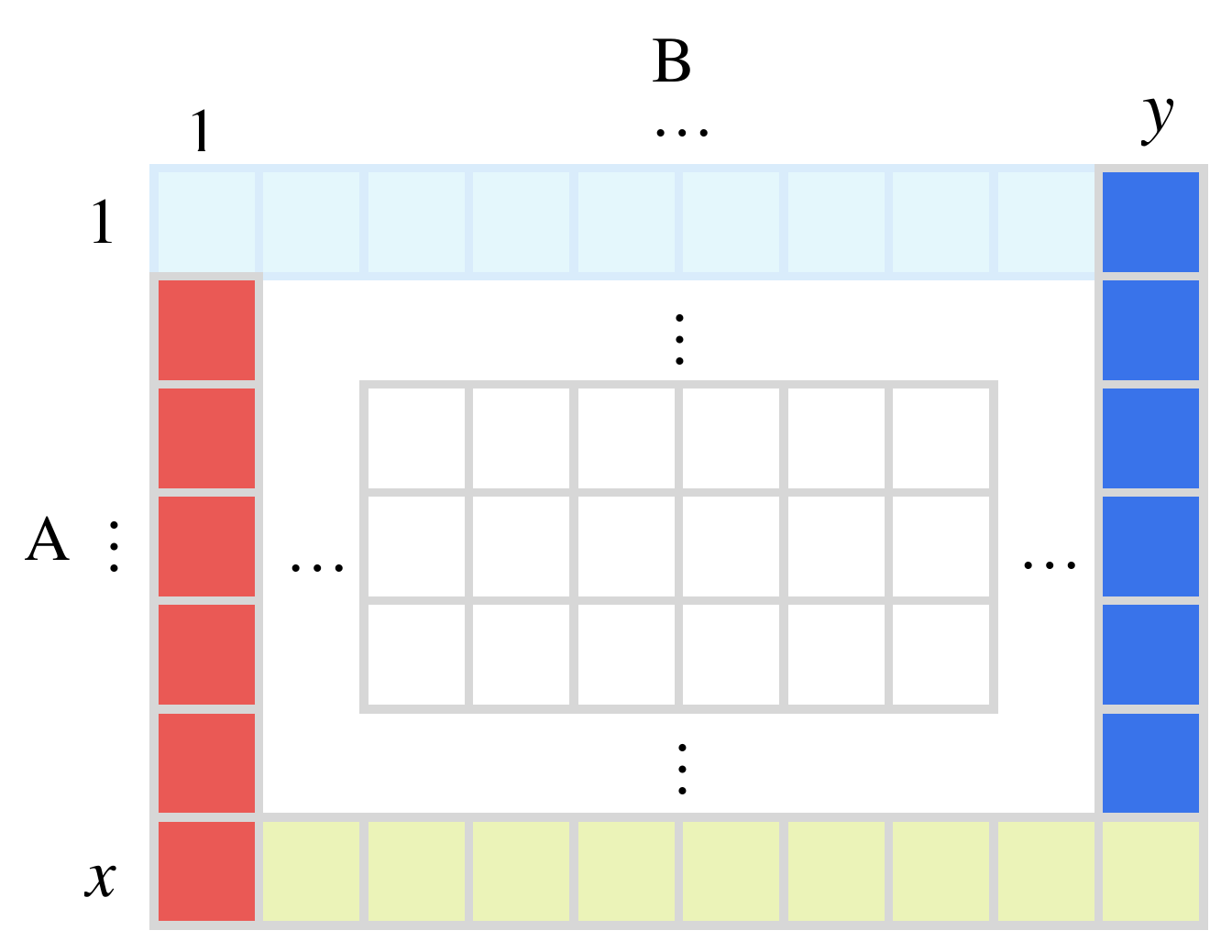}
		\caption{\label{couple_two_bi}States structure corresponding to $\{| \psi\rangle\}$  in  Theorem \ref{biparite_small_number} (or $\{| \Psi\rangle\}$  in Theorem \ref{triparite_nonlocal}). }
	\end{figure}
Motivated by the constructions of nonlocal sets of product states in Ref. \cite{Xu20a}, the following theorem is a generalized version of their results but the proof is more elegant.

\begin{theorem}\label{biparite_small_number}
Let $x,y\geq 3$ be integers and $X\in U_{FL}(x-1)$, $Y\in U_{FL}(y-1)$.  Let $\mathcal{H}_A$($\mathcal{H}_B$) be a Hilbert space of dimension $x$($y$) with  an ordered orthonormal basis  $\mathcal{A} =(|1\rangle_A,\cdots, |x\rangle_A)$($\mathcal{B} =(|1\rangle_B,\cdots, |y\rangle_B)$). The following $2(x+y)-4$ product states in $\mathcal{H}_A\otimes\mathcal{H}_B$ are locally indistinguishable (See Fig. \ref{couple_two_bi})
$$
\begin{array}{l}
|\psi_{i}\rangle:=|1\rangle_A \otimes (Y_{\mathcal{B}}^{(U)} |i\rangle_B),\\
|\psi_{y-1+j}\rangle:=(X_{\mathcal{A}}^{(U)} |j\rangle_A) \otimes   |y\rangle_B,\\
|\psi_{x+y-3+k}\rangle:=|x\rangle_A \otimes  (Y_{\mathcal{B}}^{(D)} |k\rangle_B),\\
|\psi_{x+2y-4+l}\rangle:=(X_{\mathcal{A}}^{(D)} |l\rangle_A) \otimes   |1\rangle_B,\\
\end{array}
$$
where $1\leq i\leq y-1,  1\leq j\leq x-1, 2\leq k \leq y, 2\leq l\leq x.$

\end{theorem}
\noindent {\emph{ Proof}.} Suppose Alice starts with a measurement $\{M_a^\dagger M_a\}_{a=1}^S$. The postmeasurement states should be orthogonal to each other, i.e.
$$\langle\psi_i|M_a^\dagger M_a \otimes \mathbb{I}_B|\psi_j\rangle=0,  \text{ for } i\neq j.$$ Let $M:=M_a^\dagger M_a$. Suppose its matrix representation under the ordered basis $\mathcal{A}$ is $(m_{ij})_{i,j=1}^x$. Then one finds $$M=\displaystyle \sum_{i=1}^x\sum_{j=1}^x m_{ij} |i\rangle_A \langle j|.$$

\begin{figure*}
		\includegraphics[width=0.4\textwidth,height=0.28\textwidth]{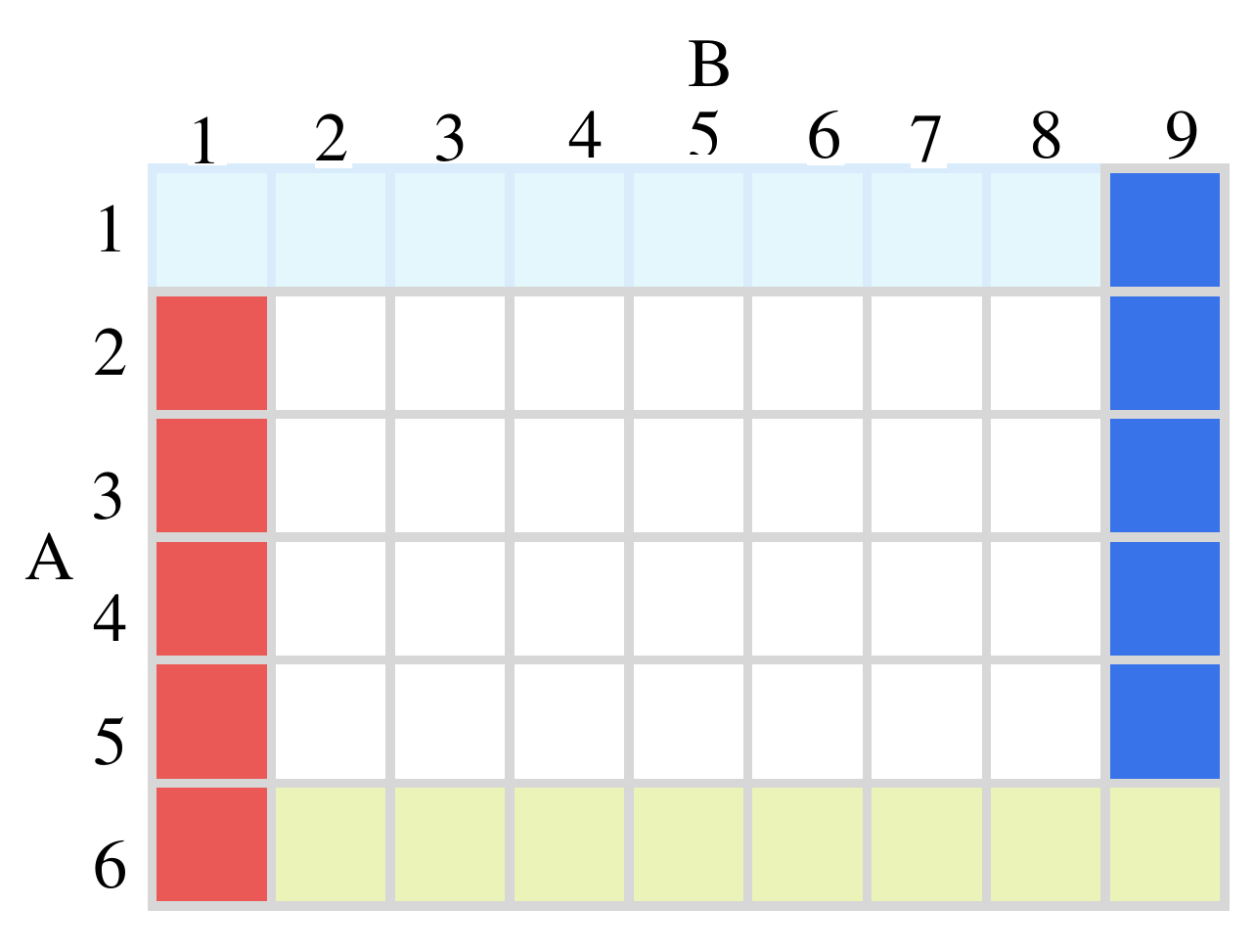}
\includegraphics[width=0.4\textwidth,height=0.28\textwidth]{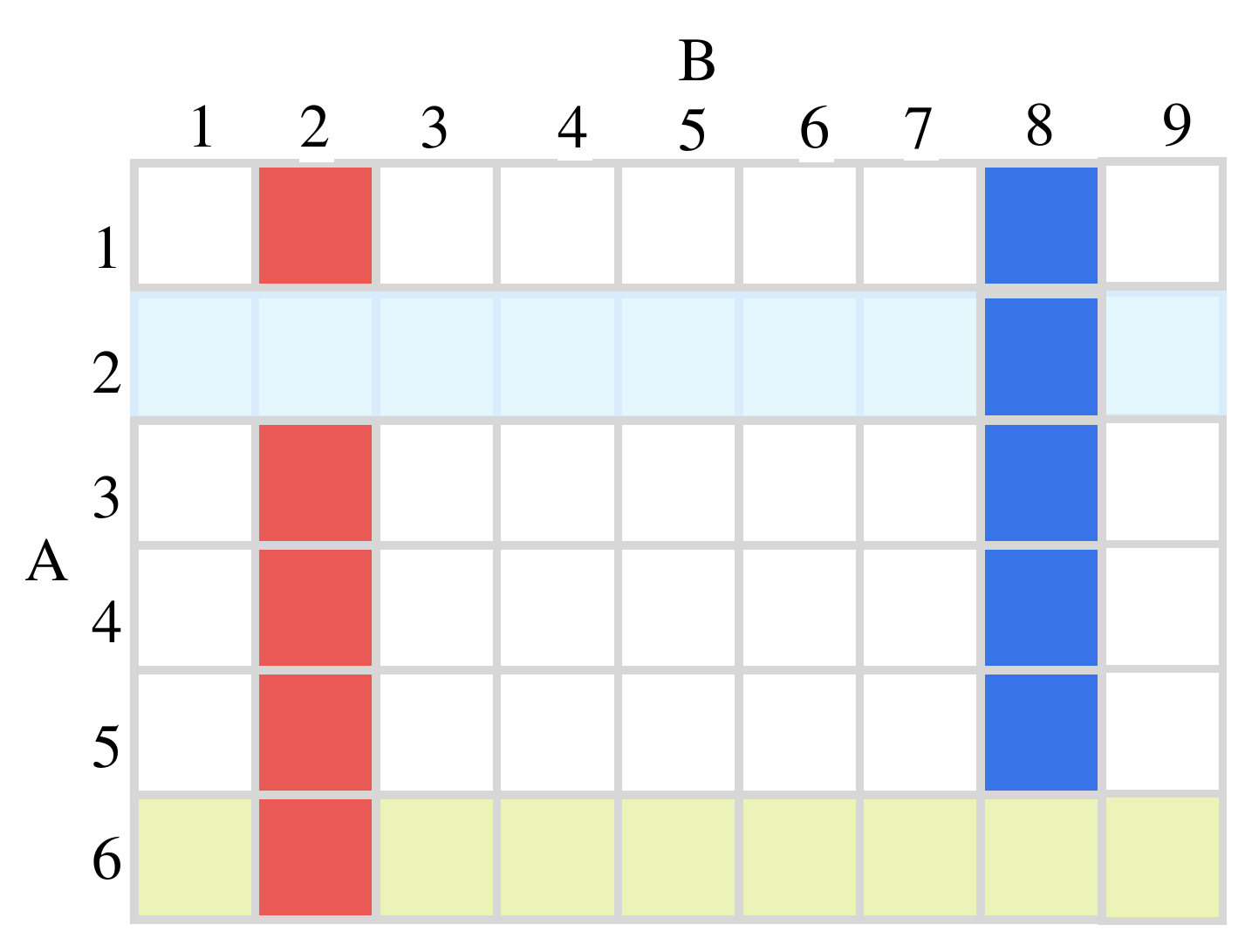}
		\caption{\label{two-compare} The left hand side draw the boundary states corresponding to the ordered bases $\mathcal{A}:=(|1\rangle_A,|2\rangle_A,|3\rangle_A,|4\rangle_A,|5\rangle_A,|6\rangle_A)$ and $\mathcal{B}:=(|1\rangle_B,|2\rangle_B,|3\rangle_B,|4\rangle_B,|5\rangle_B,|6\rangle_B,|7\rangle_B,|8\rangle_B,|9\rangle_B)$.
The right hand side presents the boundary states corresponding to the ordered bases $\mathcal{A}':=(|2\rangle_A,|1\rangle_A,|3\rangle_A,|4\rangle_A,|5\rangle_A,|6\rangle_A)$ and $\mathcal{B}':=(|2\rangle_B,|1\rangle_B,|3\rangle_B,|4\rangle_B,|5\rangle_B,|6\rangle_B,|7\rangle_B,|9\rangle_B,|8\rangle_B)$} but draws them under the ordered bases $\mathcal{A}$ and $\mathcal{B}.$
	\end{figure*}

Because $M_a\otimes \mathbb{I}_y|\psi_1\rangle$ is orthogonal to the set of states $\{M_a\otimes \mathbb{I}_y|\psi_{x+2y-4+l}\rangle \big | l=2,3,\cdots, x\}$, we have the following equations
$${}_{A}\langle 1| M X_{\mathcal{A}}^{(D)} |l\rangle_A=0,  l=2,3,\cdots, x.$$

These equalities are equivalent to the matrix equality $[m_{12},m_{13},\cdots,m_{1x}]X=[0,0,\cdots,0]$. As $X$ is invertible, we have $[m_{12},m_{13},\cdots,m_{1x}]=[0,0,\cdots,0]$. As $M$ is Hermitian, we also have $[m_{21},m_{31},\cdots,m_{x1}]=[0,0,\cdots,0]$.

Because $M_a\otimes \mathbb{I}_y |\psi_{x+y-3+2}\rangle$ is orthogonal to the set of states $\{M_a\otimes \mathbb{I}_y |\psi_{y-1+j}\rangle \big | j=1,2,\cdots, x-1\}$, we have the following equations
$${}_{A}\langle x| M X_{\mathcal{A}}^{(U)}|j\rangle_A=0,  j=1,2,\cdots, y-1.$$  These equalities are equivalent to the matrix equality $[m_{x1},m_{x2},\cdots,m_{x(x-1)}]X=[0,0,\cdots,0]$. As $X$ is invertible, we have $[m_{x1},m_{x2},\cdots,m_{x(x-1)}]=[0,0,\cdots,0]$. As $M$ is Hermitian, we also have $[m_{1x},m_{2x},\cdots,m_{(x-1)x}]=[0,0,\cdots,0]$.

Because the set of states $\{M_a\otimes \mathbb{I}_y |\psi_{y-1+j}\rangle \big | j=1,2,\cdots, x-1\}$ are pairwise orthogonal to each other, we have the following equations
$${}_A\langle j_1|{X_{\mathcal{A}}^{(U)}}^\dagger M X_{\mathcal{A}}^{(U)} |j_2\rangle_A =0,  \text{ for } 1\leq j_1\neq j_2\leq x-1. $$
If we define $M^{(u)}:=(m_{ij})_{i,j=1}^{x-1}$, the above equalities are  equivalent to  $X^\dagger M^{(u)} X=\text{diag}(\alpha_1,\cdots,\alpha_{x-1})$. Since $X$ is a unitary matrix, we have $$M^{(u)} X=X\text{diag}(\alpha_1,\cdots,\alpha_{x-1}).$$ In the following, we will compare the first row of the matrices at both hand sides of the above equality. Suppose that $X=(X_{ij})_{i,j=1}^{x-1}$.
Then the first row of $M^{(u)}X$ is $[m_{11}X_{11},m_{11}X_{12},\cdots, m_{11}X_{1(x-1)}]$. Meanwhile, the first row of $X\text{diag}(\alpha_1,\cdots,\alpha_{x-1})$ is $[\alpha_1X_{11},\alpha_2X_{12},\cdots, \alpha_{x-1}X_{1(x-1)}].$
Comparing these two vectors, one can derive
$[\alpha_1,\alpha_2,\cdots,\alpha_{x-1}]=[m_{11},m_{11},\cdots,m_{11}]$ as $X_{11},X_{12},\cdots,X_{1(x-1)}$  are all nonzero. Hence $$M^{(u)}=X\text{diag}(\alpha_1,\cdots,\alpha_{x-1})X^\dagger= m_{11}\mathbb{I}_{x-1}.$$

 Using the orthogonal relations among the states in  $\{M_a\otimes \mathbb{I}_y |\psi_{x+2y-4+l}\rangle \big | l=2,3,\cdots, x\}$, we have
$${}_A\langle l_1|{X_{\mathcal{A}}^{(D)}}^\dagger M X_{\mathcal{A}}^{(D)} |l_2\rangle_A =0,  \text{ for } 2\leq l_1\neq l_2\leq x. $$
If we define $M^{(d)}:=(m_{ij})_{i,j=2}^x$, the above equalities are  equivalent to  $X^\dagger M^{(d)} X=\text{diag}(\beta_2,\cdots,\beta_{x})$. In a similar way (but here we should consider the last row instead of the first row), we can get $
M^{(d)}=m_{xx}\mathbb{I}_{x-1}. $

 Therefore, the Hermitian matrix $M$ is of the form $m_{11} \mathbb{I}_x$. That is, Alice can only start with trivial measurement.  By the symmetry of the constructed states, Bob can only start with trivial measurement.\qed

 \vskip 5pt

 \noindent {\bf Remark:} Notice that the product states we constructed in Theorem \ref{biparite_small_number} are spanned by the ``boundary states"  with respect to the ordered bases $\mathcal{A}$ and $\mathcal{B}$ (the   outermost layer of a rectangle under the  ordered bases $\mathcal{A}$ and $\mathcal{B}$). One finds that
$ \text{span}_\mathbb{C}\{|\psi_1\rangle,\cdots, |\psi_{2x+2y-4}\rangle\}$ is equal to $$\text{span}_{\mathbb{C}}\{\mathcal{A}_i\otimes\mathcal{B}_j | i\in\{1,x\} \text{ or } j\in \{1,y\}\}$$
where $\mathcal{A}_i$($\mathcal{B}_j$) is the $i$-th ($j$-th) element of the $x$($y$)-tuples $\mathcal{A}$ ($\mathcal{B}$) (see Fig. \ref{two-compare}).


\section{Constructing genuinely nonlocal set from known ones}\label{third}
As any set of  orthogonal product states in $\mathbb{C}^2\otimes\mathbb{C}^d$ is locally distinguishable \cite{DiVincenzo03}, a necessary condition for an orthogonal set of fully product states in $\bigotimes_{i=1}^L\mathbb{C}^{d_i}$ to be genuinely nonlocal is $d_i\geq 3$ for all $i$. In this section, we show that there always exists some genuinely nonlocal set of fully product states in   $\bigotimes_{i=1}^L\mathbb{C}^{d_i}$ if the previous necessary condition is fulfilled.

\begin{theorem}\label{MainTheorem}
    Let $L\geq 3$  and $d_i\geq 3 \ (i=1,2,\cdots, L)$ be integers. Then there always exists an orthogonal set of fully product states in   $\otimes_{i=1}^L\mathbb{C}^{d_i}$ that is genuinely nonlocal.
\end{theorem}

This conclusion can be derived from  Theorem \ref{biparite_small_number}, Theorem \ref{triparite_nonlocal}, Proposition \ref{Strong_Product_four} of this paper and  the genuinely nonlocal set of $\mathbb{C}^3\otimes\mathbb{C}^3\otimes\mathbb{C}^3$ constructed in Ref. \cite{Rout1909}.

\vskip 5pt

Notice that if $S=\{|\phi_k\rangle_A|\theta_k\rangle_B\}_{k=1}^N$ is $A|B$ locally indistinguishable,  then $\mathcal{S}=\{|\phi_k\rangle_A|\theta_k\rangle_B|\varphi\rangle_{A_1}|\vartheta\rangle_{B_1}\}_{k=1}^N$ is also $AA_1|BB_1$ locally indistinguishable. Otherwise, in the first setting,  Alice and  Bob can   prepare the ancillar qudit states as $|\varphi\rangle_{A_1}$, $|\vartheta\rangle_{B_1}$  respectively on themselves  side and using the latter distinguish strategy to locally distinguish the set of states in $S$.   Moreover, we have the following observation(see also in Ref. \cite{Rout1910})

\begin{observation}\label{obser1}
		Let $S=\{\ket{\Psi_k}_{AB} \}_{k=1}^{N}$ be a   nonlocal product set shared between Alice and Bob. Consider the set $\mathcal{S}:=\{\ket{\Psi_k}_{AB}\otimes\ket{\Phi_0}_{A_1 \cdots A_m}\otimes\ket{\Theta_0}_{B_1 \cdots B_n}\}_{k=1}^{N}$, where $\ket{\Phi_0}_{A_1 \cdots A_m}$  and $\ket{\Theta_0}_{B_1 \cdots B_n}$ are some fully product states with some of the subsystems $\{A_i\}_{i=1}^m$ and $\{B_j\}_{j=1}^n$ being  in possession of some parties. The resulting set $\mathcal{S}$ is also nonlocal between $\mathcal{H}_A\otimes (\otimes_{i=1}^m\mathcal{H}_{A_i})$ and $\mathcal{H}_B\otimes (\otimes_{j=1}^n\mathcal{H}_{B_j})$.
	\end{observation}

With this observation, we show how to use the special structure of nonlocal sets  in Theorem \ref{biparite_small_number} to construct genuinely nonlocal set of product states in tripartite systems.

	\begin{figure}[h]
		\includegraphics[width=0.44\textwidth,height=0.33\textwidth]{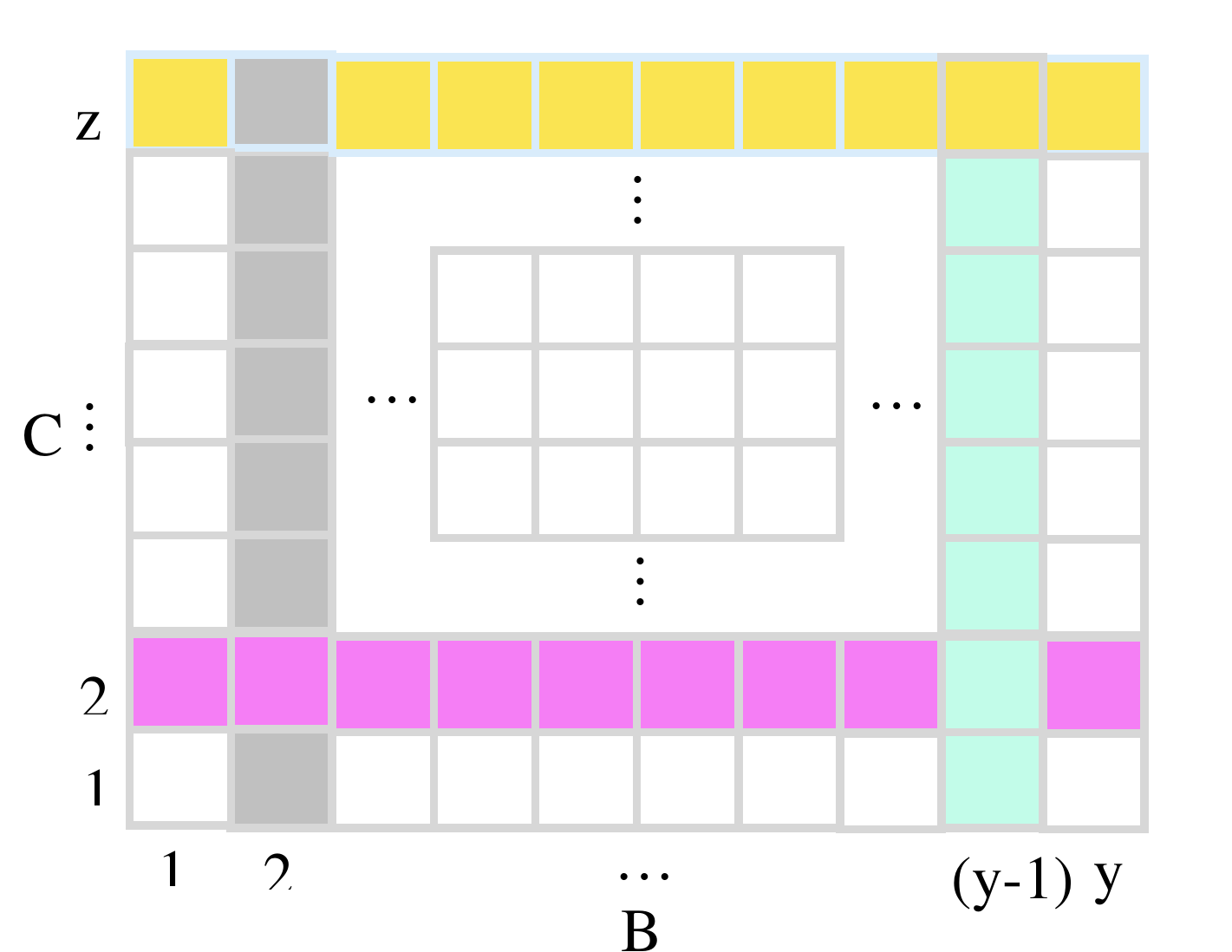}
		\caption{\label{tripart_xy}States structure  corresponding to $\{| \Phi\rangle\}$ in the Theorem \ref{triparite_nonlocal}. }
	\end{figure}

\begin{theorem}\label{triparite_nonlocal}
Let $x,z\geq 3,y\geq 4$ be integers and $X$, $Y$, $Z$  belong to $U_{FL}(x-1)$, $U_{FL}(y-1)$ and $U_{FL}(z-1)$ respectively. Let $\mathcal{H}_A$, $\mathcal{H}_B$, $\mathcal{H}_C$ be   Hilbert spaces of dimension $x,y,z$ respectively. Suppose that   $\mathcal{A} =(|1\rangle_A,\cdots, |x\rangle_A)$,  $\mathcal{B} =(|1\rangle_B,\cdots, |y\rangle_B)$, and $\mathcal{C} =(|1\rangle_C,\cdots, |z\rangle_C)$ are ordered orthonormal bases with respect to $\mathcal{H}_A$, $\mathcal{H}_B$, $\mathcal{H}_C$.  The following $2x+4y+2z-8$ product states in $\mathbb{C}^x\otimes\mathbb{C}^y\otimes \mathbb{C}^z$ are pairwise orthogonal and they form a set of product states which is genuinely nonlocal (see Fig. \ref{couple_two_bi})
$$
\begin{array}{l}
|\Psi_{i}\rangle:=|1\rangle_A\otimes  (Y_\mathcal{B}^{(U)} |i\rangle_B)\otimes |1\rangle_C,\\
|\Psi_{y-1+j}\rangle:=(X_\mathcal{A}^{(U)}|j\rangle_A)\otimes    |y\rangle_B\otimes |1\rangle_C,\\
|\Psi_{x+y-3+k}\rangle:=|x\rangle_A \otimes  (Y_\mathcal{B}^{(D)}|k\rangle_B)\otimes |1\rangle_C,\\
|\Psi_{x+2y-4+l}\rangle:=(X_\mathcal{A}^{(D)}|l\rangle_A)\otimes    |1\rangle_B\otimes |1\rangle_C,\\
\end{array}
$$
where $1\leq i\leq y-1,  1\leq j\leq x-1, 2\leq k \leq y, 2\leq l\leq x$
and (see Fig. \ref{tripart_xy})
$$
\begin{array}{l}
|\Phi_{i}\rangle:=|2\rangle_A\otimes|1'\rangle_B\otimes (Z_{\mathcal{C}'}^{(U)}|i'\rangle_C),\\
|\Phi_{z-1+j}\rangle:=|2\rangle_A\otimes(Y_{\mathcal{B}'}^{(U)})|j'\rangle_B \otimes  |z'\rangle_C,\\
|\Phi_{y+z-3+k}\rangle:=|2\rangle_A\otimes|y'\rangle_B\otimes  (Z_{\mathcal{C}'}^{(D)}|k'\rangle_C),\\
|\Phi_{y+2z-4+l}\rangle:=|2\rangle_A\otimes(Y_{\mathcal{B}'}^{(D)}|l'\rangle_B)\otimes   |1'\rangle_C.\\
\end{array}
$$
where $1\leq i\leq z-1,  1\leq j\leq y-1, 2\leq k \leq z, 2\leq l\leq y
$. Here the ordered bases $\mathcal{B}'$ and $\mathcal{C}' $ are defined as follows
$$\begin{array}{l}
\mathcal{B}':=(|1'\rangle_B,|2'\rangle_B,\cdots,|(y-1)'\rangle_B,|y'\rangle_B),\\
\mathcal{C}':=(|1'\rangle_C,|2'\rangle_C,\cdots, |z'\rangle_C)
\end{array}
$$
 where $|1'\rangle_B=|2\rangle_B$,$|2'\rangle_B=|1\rangle_B$, $|y-1'\rangle_B=|y\rangle_B$, $|y'\rangle_B=|y-1\rangle_B$, $|j'\rangle_B=|j\rangle_B$  for $3\leq j\leq y-1$ and $|1'\rangle_C=|2\rangle_C$,$|2'\rangle_C=|1\rangle_C$, $|k'\rangle_C=|k\rangle_C$  for $3\leq j\leq z$.

\end{theorem}

	\begin{figure}[h]
		\includegraphics[width=0.48\textwidth,height=0.34\textwidth]{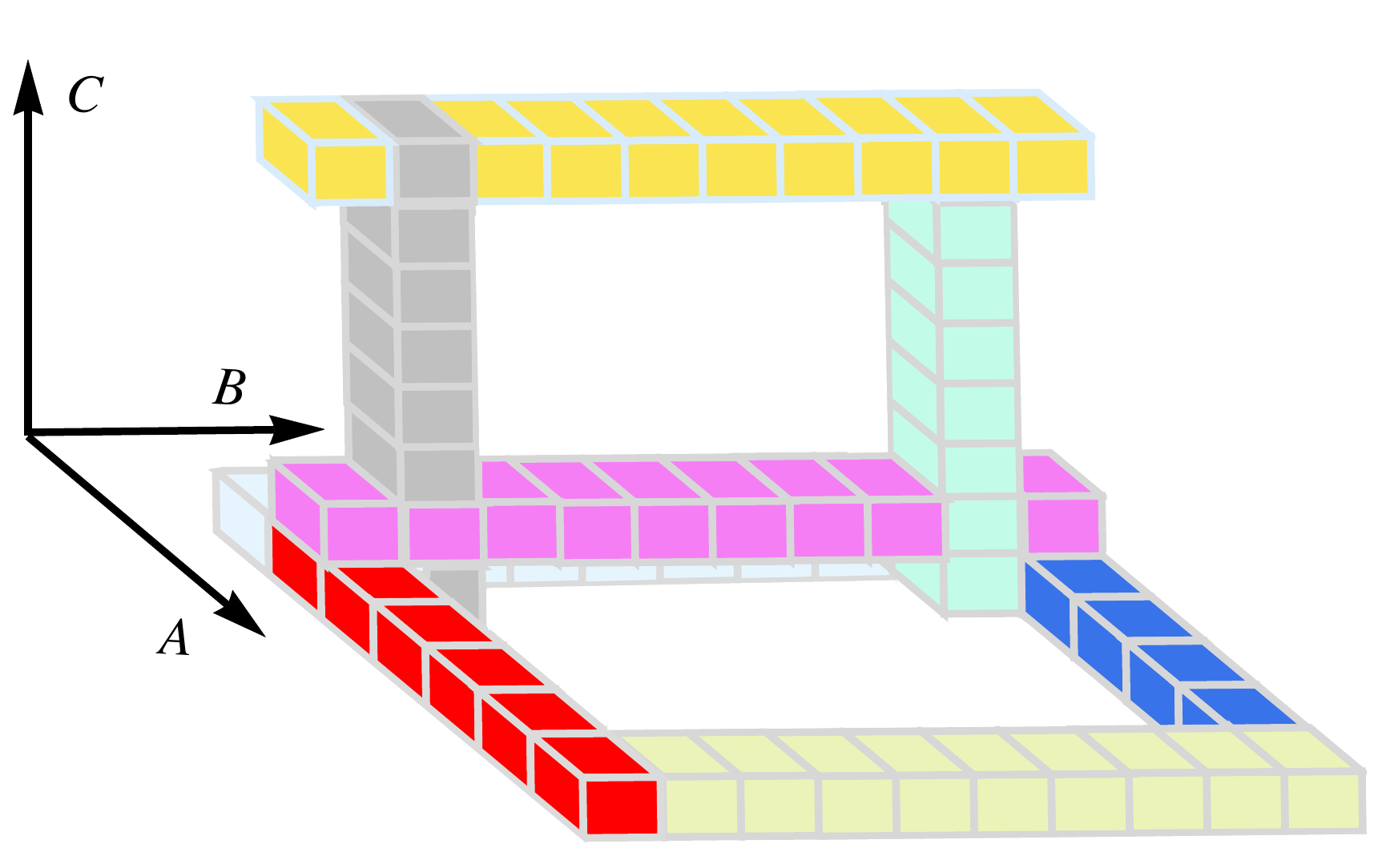}
		\caption{\label{tripart_cubic}This is a schematic diagram for the states in Theorem  \ref{triparite_nonlocal}.}
	\end{figure}

	\noindent \emph{Proof.} We notice that
$ \text{span}_\mathbb{C}\{|\Psi_1\rangle,|\Psi_2\rangle\cdots, |\Psi_{2x+2y-4}\rangle\}$ is equal to the linear space spanned by
 $$\mathcal{S}_{\Psi}:=\{|i\rangle_A|j\rangle_B|1\rangle_C \big | \ \  i\in\{1,x\} \text{ or } j\in \{1,y\}\}$$ while $ \text{span}_\mathbb{C}\{|\Phi_1\rangle,|\Phi_2\rangle,\cdots, |\Phi_{2y+2z-4}\rangle\}$ is equal to the linear space spanned by
 $$\begin{array}{l}
 \mathcal{S}_{\Phi}:=\{|2\rangle_A|j'\rangle_B|k'\rangle_C \big | \ \  j\in\{1,y\} \text{ or } k\in \{1,z\}\}\\
 =\{|2\rangle_A|j\rangle_B|k\rangle_C \big | \ \  j\in\{2,y-1\} \text{ or } k\in \{2,z\}\}.
 \end{array}$$
 As $\mathcal{S}_{\Psi}\cap \mathcal{S}_{\Phi}=\emptyset$ and $\mathcal{S}_{\Psi},\mathcal{S}_{\Phi}\subseteq \{|i\rangle_A|j\rangle_B|k\rangle_C \big | \ 1\leq i\leq x,1\leq j\leq y,1\leq k\leq z\}$ which is a orthonormal basis of $\mathcal{H}_A\otimes \mathcal{H}_B\otimes\mathcal{H}_C$, we have  $\langle \Psi_u |\Phi_v\rangle=0$ for integers $u,v$ with  $1\leq u\leq 2x+2y-4,1\leq v\leq 2y+2z-4.$ Therefore, the states in $\{|\Psi_u\rangle\}_{u=1}^{2x+2y-4} \cup \{|\Phi_v\rangle\}_{v=1}^{2y+2z-4}$  are pairwise orthogonal ( Fig. \ref{tripart_cubic} is more intuitive for the orthogonality).

To prove that the set of states we construct are genuinely nonlocal.  There are only three ways to separate $ABC$ into two sets. That is, $A|BC, B|CA,C|AB$. By  Theorem \ref{biparite_small_number} and Observation \ref{obser1},  the set $\{|\Psi_u\rangle\}_{u=1}^{2x+2y-4}$ is locally indistinguishable as the partitions $A|BC$ and $B|CA$. And the set $\{|\Phi_v\rangle\}_{v=1}^{2y+2z-4}$ is locally indistinguishable as the partitions $B|CA$ and $C|AB$.  Hence the given set is genuinely nonlocal. \qed

\vskip 6pt
  	In the following, we begin to strengthen the results in \cite{Zhang17} where they constructed locally indistinguishable multipartite product states from known bipartite ones. The following two propositions enhance  their results to genuine nonlocality settings.

  	\begin{figure}[h]
	\includegraphics[width=0.45\textwidth,height=0.24\textwidth]{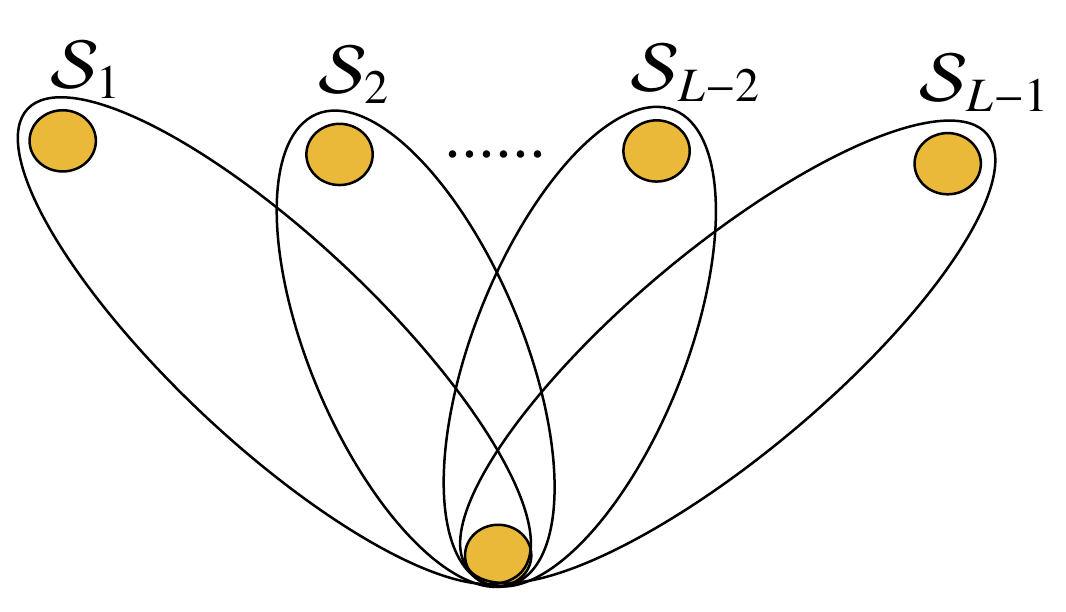}
	\caption{\label{couple_one_to_l}States structure corresponding to the Proposition \ref{Strong_Product_four}.  }
\end{figure}
  		\begin{proposition}\label{Strong_Product_four}
		Let $L\geq 4$ be an integer and $d_i\geq 3$ for all $1\leq i\leq L$.  Let $S_i=\{|\psi_j^{(i)}\rangle |\phi_j^{(i)}\rangle\}_{j=1}^{n_i}\subseteq\mathbb{C}^{d_1}\otimes\mathbb{C}^{d_{i+1}}$ be sets of product states that are locally indistinguishable for $i=1,2,\cdots,L-1$. Then the  union of the following sets    (See Fig. \ref{couple_one_to_l})
	\begin{equation}\label{states_all_to_L1}
		\begin{array}{lllll}
		\mathcal{S}_1=\{|\psi_j^{(1)}\rangle|\phi_j^{(1)}\rangle|1\rangle|1\rangle|1\rangle\cdots|2\rangle\}_{j=1}^{n_1}, \\[2mm]
		\mathcal{S}_2=\{|\psi_j^{(2)}\rangle|2\rangle|\phi_j^{(2)}\rangle|1\rangle|1\rangle\cdots|1\rangle\}_{j=1}^{n_2},\\[2mm]
		\mathcal{S}_3=\{|\psi_j^{(3)}\rangle|1\rangle|2\rangle|\phi_j^{(3)}\rangle|1\rangle\cdots|1\rangle\}_{j=1}^{n_3},\\[2mm]
		\ \ \ \ \ \ \ \ \ \ \ \vdots\\[2mm]
		\mathcal{S}_{L-1}=\{|\psi_j^{(L-1)}\rangle|1\rangle|1\rangle\cdots|1\rangle|2\rangle|\phi_j^{(L-1)}\rangle\}_{j=1}^{n_{L-1}}
		\end{array}
		\end{equation}
		is also a genuinely nonlocal set of  product states in $\mathbb{C}^{d_1}\otimes\mathbb{C}^{d_2}\otimes\cdots\otimes\mathbb{C}^{d_L}$.
	\end{proposition}

\emph{Proof.}   To distinguish the states of $\mathcal{S}_1$, by Observation \ref{obser1},  the first two parties must come together and perform a global measurement. Similarly, to distinguish the states of $\mathcal{S}_i$, the $1$-th and ($i+1$)-th parties must come together and perform a global measurement.
	
	Therefore, all the parties must come together to distinguish all the states in Eq. (\ref{states_all_to_L1}). Hence such a set of states is genuinely nonlocal. \qed

	\begin{figure}[h]
		\includegraphics[width=0.5\textwidth,height=0.14\textwidth]{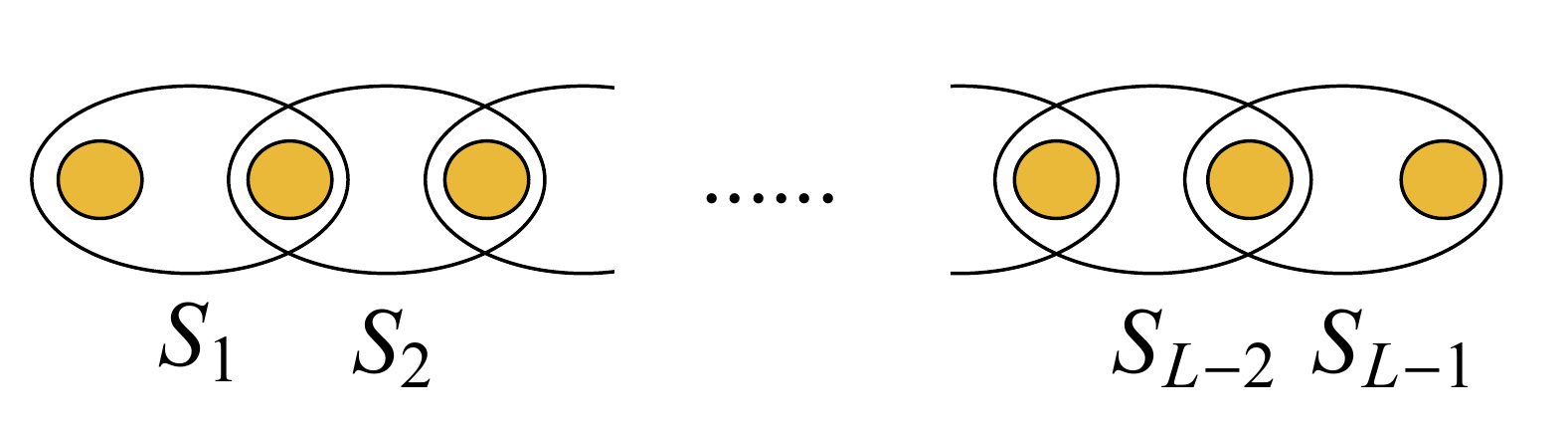}
		\caption{\label{couple_two}States structure corresponding to the Proposition \ref{Strong_Product_five}. }
	\end{figure}
	\begin{proposition}\label{Strong_Product_five}Let $L\geq 5$ be an integer and $d_i\geq 3$ for all $1\leq i\leq L$.  Let $S_i:=\{|\psi_j^{(i)}\rangle |\phi_j^{(i)}\rangle\}_{j=1}^{n_i}\subseteq\mathbb{C}^{d_i}\otimes\mathbb{C}^{d_{i+1}}$ be sets of product states that are locally indistinguishable for $i=1,2,\cdots,L-1$. Then the union of the following sets  (See Fig. \ref{couple_two})
	\begin{equation}\label{states_two_to_L1}
\begin{array}{lllll}
\mathcal{S}_1=\{|\psi_j^{(1)}\rangle|\phi_j^{(1)}\rangle|2\rangle|1\rangle|1\rangle|1\rangle\cdots|1\rangle\}_{j=1}^{n_1}, \\[2mm]
\mathcal{S}_2=\{|1\rangle|\psi_j^{(2)}\rangle|\phi_j^{(2)}\rangle|2\rangle|1\rangle|1\rangle\cdots|1\rangle\}_{j=1}^{n_2},\\[2mm]
\mathcal{S}_3=\{|1\rangle|1\rangle|\psi_j^{(3)}\rangle|\phi_j^{(3)}\rangle|2\rangle|1\rangle\cdots|1\rangle\}_{j=1}^{n_3},\\[2mm]
\ \ \ \ \ \ \ \ \ \ \ \vdots\\[2mm]
\mathcal{S}_{L-1}=\{|2\rangle|1\rangle|1\rangle|1\rangle\cdots|1\rangle|\psi_j^{(L-1)}\rangle|\phi_j^{(L-1)}\rangle\}_{j=1}^{n_{L-1}}
\end{array}
	\end{equation}
	is also a genuinely nonlocal set of  product states in $\mathbb{C}^{d_1}\otimes\mathbb{C}^{d_2}\otimes\cdots\otimes\mathbb{C}^{d_L}$.

		\end{proposition}

To construct multipartite genuinely  nonlocal sets, instead of using bipartite nonlocal product states, we can also start with some known genuinely nonlocal sets of product states in tripartite systems.
	
\begin{figure}[h]
	\includegraphics[width=0.45\textwidth,height=0.24\textwidth]{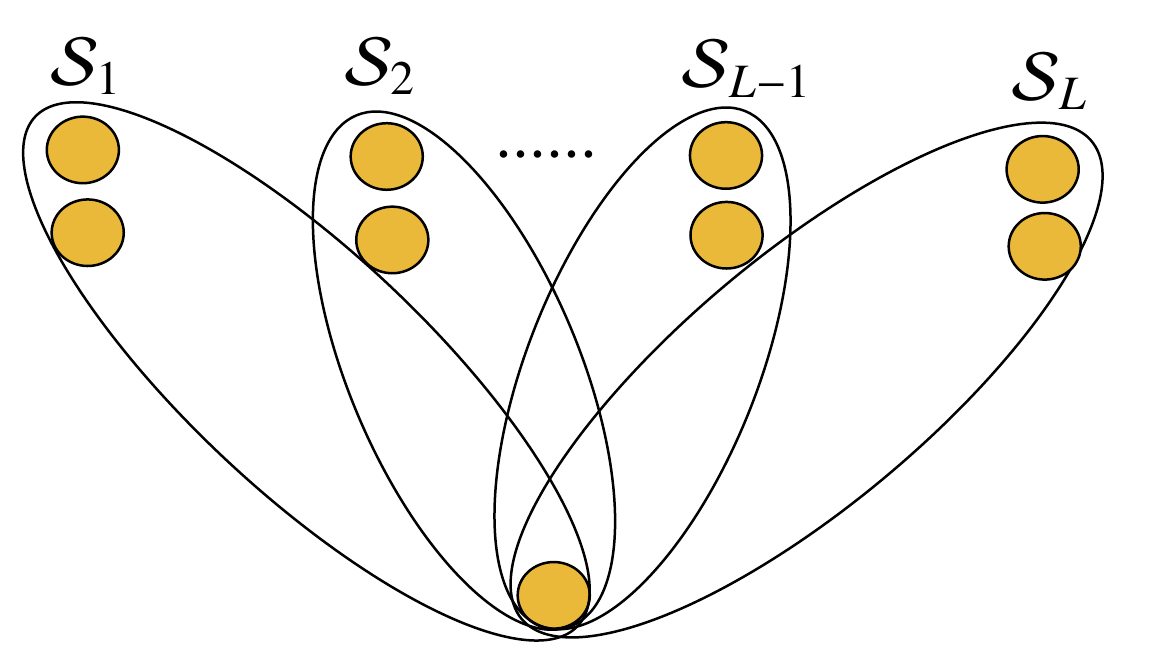}
	\caption{\label{couple_one_with_twoall}States structure corresponding to the Proposition \ref{Strong_Three_To_Mul}.  }
\end{figure}

	\begin{proposition}\label{Strong_Three_To_Mul}
	Let $L\geq 3$ be an integer.	Let $\{|\psi_j\rangle |\phi_j\rangle|\chi_j\rangle\}_{j=1}^{n}\subseteq\mathbb{C}^{3}\otimes\mathbb{C}^{3}\otimes\mathbb{C}^{3}$  be a set of product states that is genuinely nonlocal. Then the  union of the following sets   (See Fig. \ref{couple_one_with_twoall})
		$$
		\begin{array}{lllll}
	\mathcal{S}_1=\{	|\psi_j\rangle|\phi_j\rangle|\chi_j\rangle |1\rangle|1\rangle|1\rangle\cdots|1\rangle|2\rangle|2\rangle\}_{j=1}^{n},  \\[2mm]
	\mathcal{S}_2=\{	|\psi_j\rangle|2\rangle|2\rangle|\phi_j\rangle|\chi_j\rangle|1\rangle \cdots|1\rangle|1\rangle|1\rangle\}_{j=1}^{n}, \\[2mm]
	\mathcal{S}_3=\{	|\psi_j\rangle|1\rangle|1\rangle|2\rangle|2\rangle|\phi_j\rangle|\chi_j\rangle|1\rangle\cdots|1\rangle\}_{j=1}^{n}, \\[2mm]
		\ \ \ \ \ \ \ \ \ \ \ \ \ \ \vdots\\[2mm]
	\mathcal{S}_L=\{	|\psi_j\rangle|1\rangle|1\rangle|1\rangle|1\rangle\cdots|2\rangle|2\rangle|\phi_j\rangle|\chi_j\rangle\}_{j=1}^{n},
		\end{array}
		$$
		is also a genuinely nonlocal set of  product states   in $\bigotimes_{i=1}^{2L+1} \mathcal{H}_{A_i}$ where $ \mathcal{H}_{A_i}=\mathbb{C}^3$.
	\end{proposition}

In fact, the above constructions can be extended  to much more general settings. Let $L\geq 3$ be an integer and $\mathcal{P}:=\{1,2,3,\cdots,L\}$. Let $\mathcal{H}:=\otimes_{j\in\mathcal{P}}\mathcal{H}_j$ be an $L$-parties quantum system. Suppose there are $s$ proper subsets of  $\mathcal{P}$: $\mathcal{P}_1, \mathcal{P}_2,\cdots,\mathcal{P}_s$   and we denote $\overline{\mathcal{P}}_i:=\mathcal{P}\setminus \mathcal{P}_i$ for each $i$.  We make the following assumptions:
 \begin{enumerate}
   \item [{\rm (a)}] ${S}_i=\{|\Psi_j\rangle_{\mathcal{P}_i}\}_{j=1}^{n_i} $   is a genuinely nonlocal product set in $\mathcal{H}_{\mathcal{P}_i}:=\otimes_{j\in\mathcal{P}_i} \mathcal{H}_j$ for each $i\in \{1,2,\cdots,s\} $.
   \item [{\rm (b)}] There is  a fully product state $|\Phi_i\rangle_{\overline{\mathcal{P}}_i}\in \otimes_{j\in\overline{\mathcal{P}}_i}\mathcal{H}_j$ for each $i$ such that
the states in the union of the sets $\mathcal{S}_i=\{|\Psi_j\rangle_{\mathcal{P}_i}|\Phi_i\rangle_{\overline{\mathcal{P}}_i}\}_{j=1}^{n_i}$ ($1\leq i\leq s$) are mutually orthogonal to each other.
 \end{enumerate}

We use the notation $\mathfrak{P}:=(\mathcal{P}, \{\mathcal{P}_1, \mathcal{P}_2,\cdots,\mathcal{P}_s\})$.  For each $\mathfrak{P}$, We attach it  a graph $G_{\mathfrak{P}}=(V_\mathfrak{P},E_\mathfrak{P})$   defined as follows:  its vertex set is $V_\mathfrak{P}=\mathcal{P}$ and its edge set is $$E_\mathfrak{P}=\bigcup_{i=1}^s\{(u_i,v_i)| u_i,v_i\in \mathcal{P}_i \text{ and } u_i\neq v_i \}.$$
	
	\begin{theorem}\label{general_genuine_nonlocal} Under the notation and assumptions of the two paragraphs previous and suppose that $G_\mathfrak{P}$ is  connected,
		then the set $\mathcal{S}:=\cup_{i=1}^s \mathcal{S}_i$  is a genuinely nonlocal set of  product states in $\otimes_{j\in\mathcal{P}}\mathcal{H}_j$.
	\end{theorem}
\noindent \emph{ Proof.}
  Suppose not, there exist a nontrivial bipartition of $\mathcal{P}$, say  $U\ |\  V$ (both $U$ and $V$ are nonempty subset of $\mathcal{P}$), such that the set $\mathcal{S}$ is locally distinguishable when considering as a set of bipartite states in $(\otimes_{j\in{U}}\mathcal{H}_j)\bigotimes (\otimes_{j\in V}\mathcal{H}_j).$ As the connectivity of $G_\mathfrak{P}$, there must exist some edge $(u,v)\in E_\mathfrak{P}$ which connects the two sets $U$ and $V$, i.e.   $u\in U$ and $v\in V$. By the definition of $E_\mathfrak{P}$, there exist some $i$  such that $u,v\in\mathcal{P}_i$. However, the set $S_i$ is genuinely nonlocal in $\mathcal{H}_{\mathcal{P}_i}$ by the assumption (a) above. So it is locally indistinguishable for the partition
$$(U\cap \mathcal{P}_i) \ | \  (V\cap \mathcal{P}_i)$$
of $\mathcal{P}_i$ as both $U\cap \mathcal{P}_i$ and $V\cap \mathcal{P}_i$ are nonempty.
By Observation \ref{obser1}, the set $\mathcal{S}_i$ is locally indistinguishable in the bipartite system $(\otimes_{j\in{U}}\mathcal{H}_j)\bigotimes (\otimes_{j\in V}\mathcal{H}_j).$ However, $\mathcal{S}_i\subseteq \mathcal{S}$ implies that $\mathcal{S}$ must be locally indistinguishable  as bipartite states $(\otimes_{j\in{U}}\mathcal{H}_j)\bigotimes (\otimes_{j\in V}\mathcal{H}_j).$ Hence we obtain a contradiction. So the set $\mathcal{S}$ must be genuinely nonlocal.

 \qed

	\section{Conclusion and Discussion}\label{fifth}
	
	We study   a strong form of locally indistinguishable  set of fully product states called genuinely nonlocal set. We generalize the results of locally indistinguishable product states in bipartite system in Ref. \cite{Xu20a} but  provide a much more elegant proof. Based on a simple observation,   we extend  the results of Zhang \emph{et  al.} in Ref. \cite{Zhang17}   to the cases of  genuinely nonlocal sets.  Moreover, we  extend these results to a much more general setting by relating the construction of genuinely nonlocal sets with the connectivity of some graphs.  As a consequence, we can show that there always exists some genuinely nonlocal set of fully product states in   $\otimes_{i=1}^L\mathbb{C}^{d_i}$ provided $d_i\geq 3$ for all $i$.    One should note that the genuinely nonlocal set we constructed here maybe locally reducible  under the concept introduced in Ref. \cite{Halder19}. Therefore, it is interesting to find some  method to characterize the  locally irreducible settings.

	\vspace{2.5ex}
	
	\noindent{\bf Acknowledgments}\, \,   This  work  is supported  by  National  Natural  Science  Foundation  of  China  (11771419, 11875160, 11901084, 12005092,  and  U1801661), the China Postdoctoral Science Foundation (2020M681996), the  Natural  Science  Foundation  of  Guang-dong  Province  (2017B030308003),   the  Key  R$\&$D  Program of   Guangdong   province   (2018B030326001),   the   Guang-dong    Innovative    and    Entrepreneurial    Research    TeamProgram (2016ZT06D348), the Science, Technology and   Innovation   Commission   of   Shenzhen   Municipality (JCYJ20170412152620376   and   JCYJ20170817105046702 and  KYTDPT20181011104202253),   the Economy, Trade  and  Information  Commission  of  Shenzhen Municipality (201901161512),
	the Research startup funds of DGUT (GC300501-103),  the Fundamental Research Funds for the Central Universities, and Anhui Initiative in Quantum Information Technologies under Grant No. AHY150200.

\end{document}